\documentclass[mathleft
]{an}
\usepackage{graphicx}
\usepackage{times}
\def\as         {$^{\prime\prime}$}
\begin{document}

\Pagespan{1}{}
\Yearpublication{2006}%
\Yearsubmission{2005}%
\Month{11}%
\Volume{999}%
\Issue{88}%

\title{Metal jumps across sloshing cold fronts: the case of A496}

\author{S. Ghizzardi \inst{1}\fnmsep\thanks{Corresponding author:
  \email{simona@iasf-milano.inaf.it}\newline}
\and  S. De Grandi\inst{2}
\and S. Molendi\inst{1}
}
\titlerunning{Metal jumps across A496 cold fronts.}
\authorrunning{S. Ghizzardi \& S. De Grandi \& S. Molendi}
\institute{
INAF/IASF-Milano, via Bassini 15, I20133, Milano, Italy
\and 
INAF - Osservatorio Astronomico di Brera, via E. Bianchi 46, I23807 Merate (LC), Italy}

\received{2 July 2012}
\accepted{2 July 2012}
\publonline{later}


\abstract{%
Cold-fronts in cool-core clusters are thought to be induced by minor
mergers and to develop through a sloshing mechanism. While temperature
and surface-brightness jumps have been detected and measured in many
systems, a detailed characterization of the metal
abundance across the discontinuity is only available for a handful of
objects.   Within the sloshing scenario, we expect the
central cool and metal rich gas to be displaced outwards into lower
abundance regions, thus generating a metal discontinuity across the
front.  We analyzed a long (120 ksec) XMM-Newton observation of A496 to study the
metal distribution and its correlation with the cold-fronts.  
We find Fe discontinuities across the two main cold-fronts 
located ~60 kpc NNW and ~160 kpc South of the peak and a metal excess in the 
South direction.}

\maketitle

\section{Introduction}

After that Chandra discovered cold fronts (Markevitch et al.~2000), these features have been detected in 
a large number of galaxy clusters and have been found to be a common phenomenon in the cluster 
population. Indeed, the majority of clusters host at least one cold front 
(Markevitch, Vikhlinin \& Forman~2003; Ghizzardi, Rossetti and Molendi~2010, hereafter G10).

Cold fronts origin and evolution have been widely studied in the last years by several authors, 
both using X-ray observations and  numerical simulations 
and a global picture for these phenomena has been provided (see Markevitch \& Vikhlinin~2007 for a review).
However, relatively little attention has been paid to the metallicity properties of cold fronts 
(see e.g. Simionescu et al.~2010;  de Plaa et al.~2010), 
especially in the 
sloshing scenario of cool core clusters. These clusters have prominent metallicity peaks in their centers (e.g.
De Grandi \& Molendi~2002) and hence the mixing of the gas induced by the sloshing could be important. 

Sloshing cold fronts are thought to be triggered in cool core clusters by minor mergers: an infalling substructure induces 
perturbations to the underlying gravitational potential of the main cluster and consequently the cold inner gas 
is displaced from the center of the potential well, it is decoupled from the dark matter through ram pressure 
and starts to slosh.
High resolution hydrodynamical simulations have been developed to detail 
the history of the gas during the sloshing (see e.g. Ascasibar \& Markevitch~2006;
Roediger \& ZuHone~2012; ZuHone et al.~2010,~2011).
The effect of sloshing on metal distribution in relaxed clusters have been recently addressed 
in simulations tailored to reproduce the Virgo and A496 clusters (Roediger et al. 2011,~2012).
In these simulations, the sloshing of the central enriched gas creates metal abundances 
jumps across the cold fronts and redistributes the heavy elements throughout the ICM.

On the observational side, a detailed characterization of the metal abundance across the fronts
is only available for a handful of objects (Perseus: Fabian et al.~2011; Centaurus: Sanders \& Fabian~2006; 
A2204: Sanders, Fabian \& Taylor~2005,~2009).
This is a significant limitation as the sloshing mechanism may play a crucial role in carrying metals from the core 
to the outskirts and a precise characterization of the metal distribution in a sloshing ICM 
can provide information on the enrichment and metal mixing processes at work. 

\begin{figure}
\centering
 \includegraphics[width=55mm,angle=270]{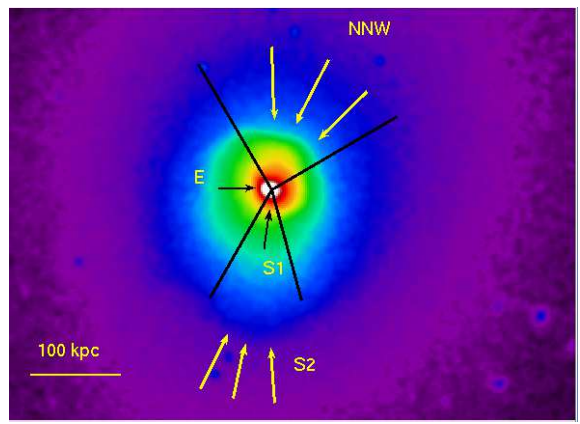}

\caption{EPIC flux map for A496.}
\label{fig:epic_fx}
\end{figure}

\begin{figure*}
\centering
\includegraphics[width=0.72\textwidth,angle=90]{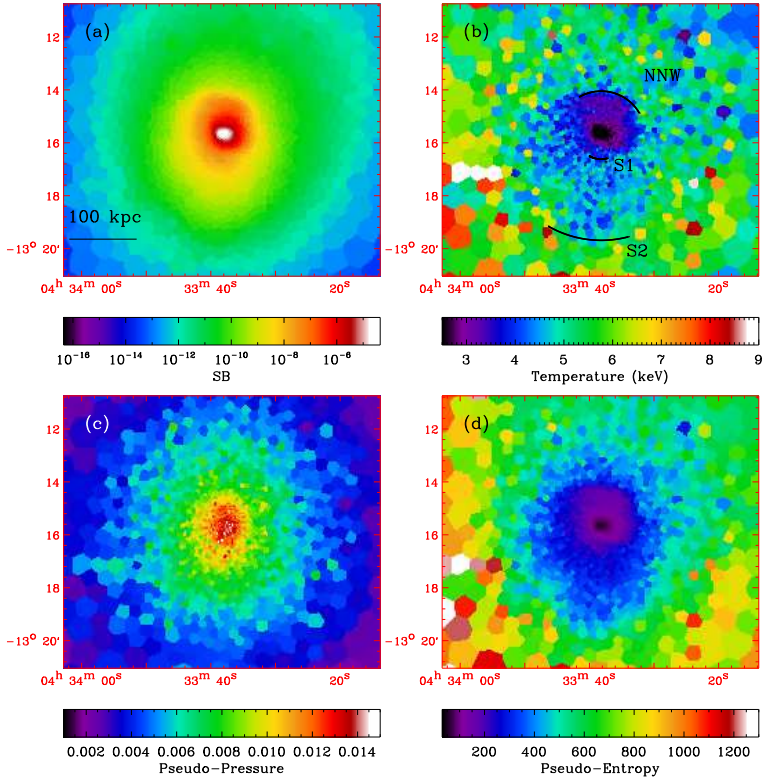}

\caption{EPIC SB image (a), temperature map (b), pseudo-pressure map (c) and pseudo-entropy map (d) for A496. Arcs 
and labels in panel (b) mark the cold fronts.}
\label{fig:mappe_wvt}
\end{figure*}
In this paper we report our results on the metal distribution in the galaxy cluster A496 obtained 
using a long  (120 ksec) XMM-Newton observation.
A496 is a good candidate to inspect the correlation between metal distribution and cold fronts as it's a bright, 
nearby, cool core cluster hosting at least four cold fronts.

\section{Results}

\subsection{Surface brightness discontinuities in A496}

XMM-Newton data for A496 have been processed in a standard way: all the 
details concerning the data reduction and the analysis procedures are provided in Ghizzardi et al. (in preparation).

To look for cold fronts in A496, we built the EPIC flux map in the 0.2-4 keV energy band, 
following the method described in G10.
The resulting flux map is shown in Fig. \ref{fig:epic_fx}. 
A visual inspection of the map clearly reveals several sharp discontinuities.
The main discontinuity is located at NNW (30$^o$-120$^o$ measured from the W direction) at a distance $\sim$ 60 kpc 
($\sim$ 100\as) from the X--ray peak. The morphology of the front appears boxy.

A surface brightness discontinuity is then observed at $\sim$ 35 kpc ($\sim$ 55\as) in the South direction (labeled 
S1 in Fig. \ref{fig:epic_fx}).
Both these discontinuities had also been detected by Dupke \& White (2003) using Chandra data.
The high spatial resolution of Chandra allowed Dupke \& White (2003) to detect another innermost cold front 
($\sim$16 kpc; labeled E in Fig. \ref{fig:epic_fx}) East of the center.
This cold front is too close to the center to be resolved by the
XMM-Newton instruments.

In the  South direction (240$^o$-285$^o$), another outer discontinuity  is detected at  
$\sim$ 160 kpc ($\sim$ 240\as), marked S2 in Fig. \ref{fig:epic_fx} (see also Tanaka et al.~2006).
This front could not be detected in Chandra maps as it lies almost at the edge of the ACIS-S3 chip.

\subsection{Thermodynamical maps and A496 cold fronts.}

To assess the nature of the discontinuities we checked the temperature behavior across the fronts.
To this aim, following the Rossetti et al. (2007) method, we built the thermodynamical maps 
(surface brightness, temperature, pseudo-pressure and pseudo-entropy) for A496 which are shown 
in Fig. \ref{fig:mappe_wvt}.

\begin{figure}
\centering
\vspace{-5truemm}
{\includegraphics[width=0.25\textwidth]{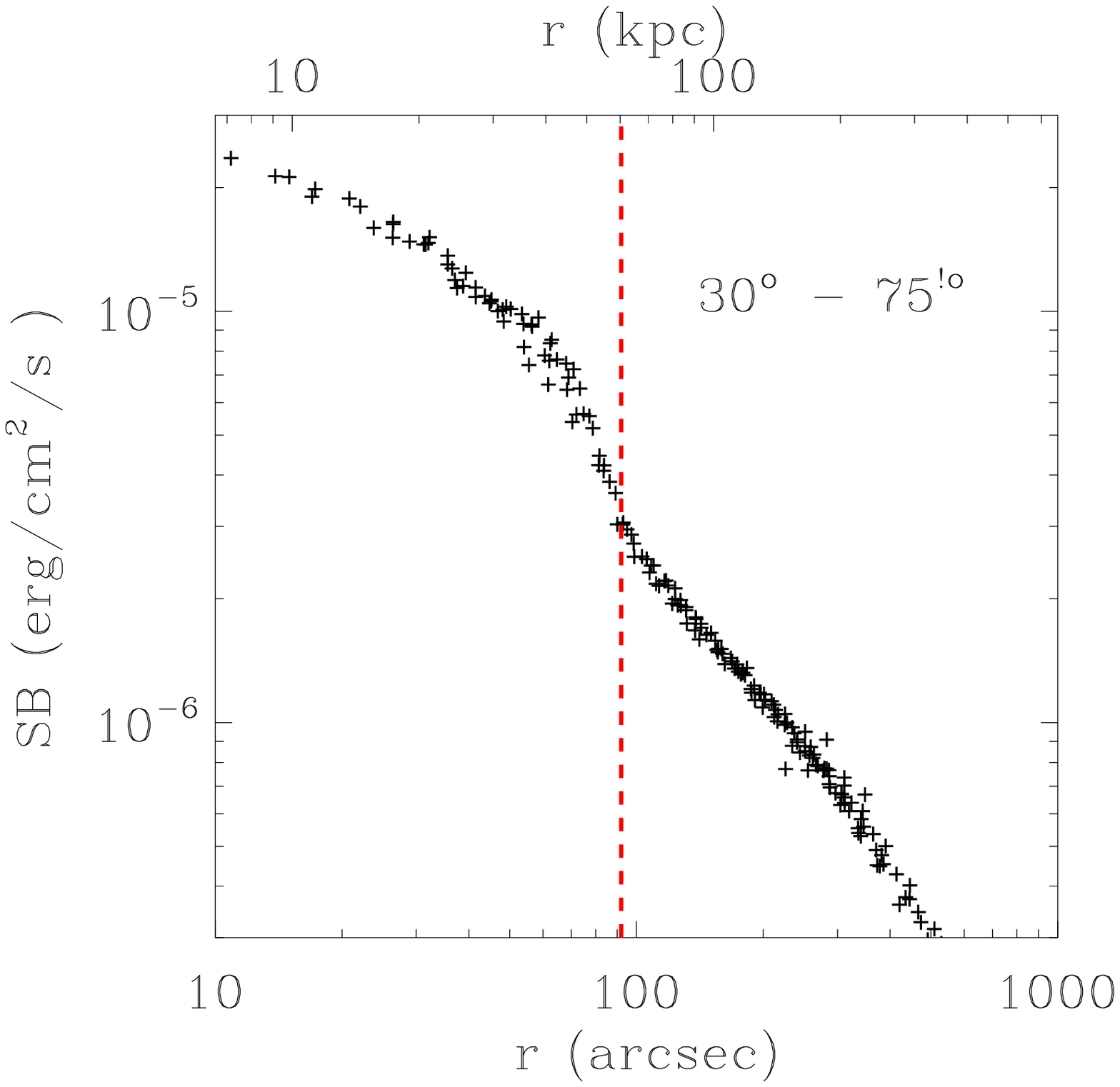}}
\hspace{-7truemm}
{\includegraphics[width=0.25\textwidth]{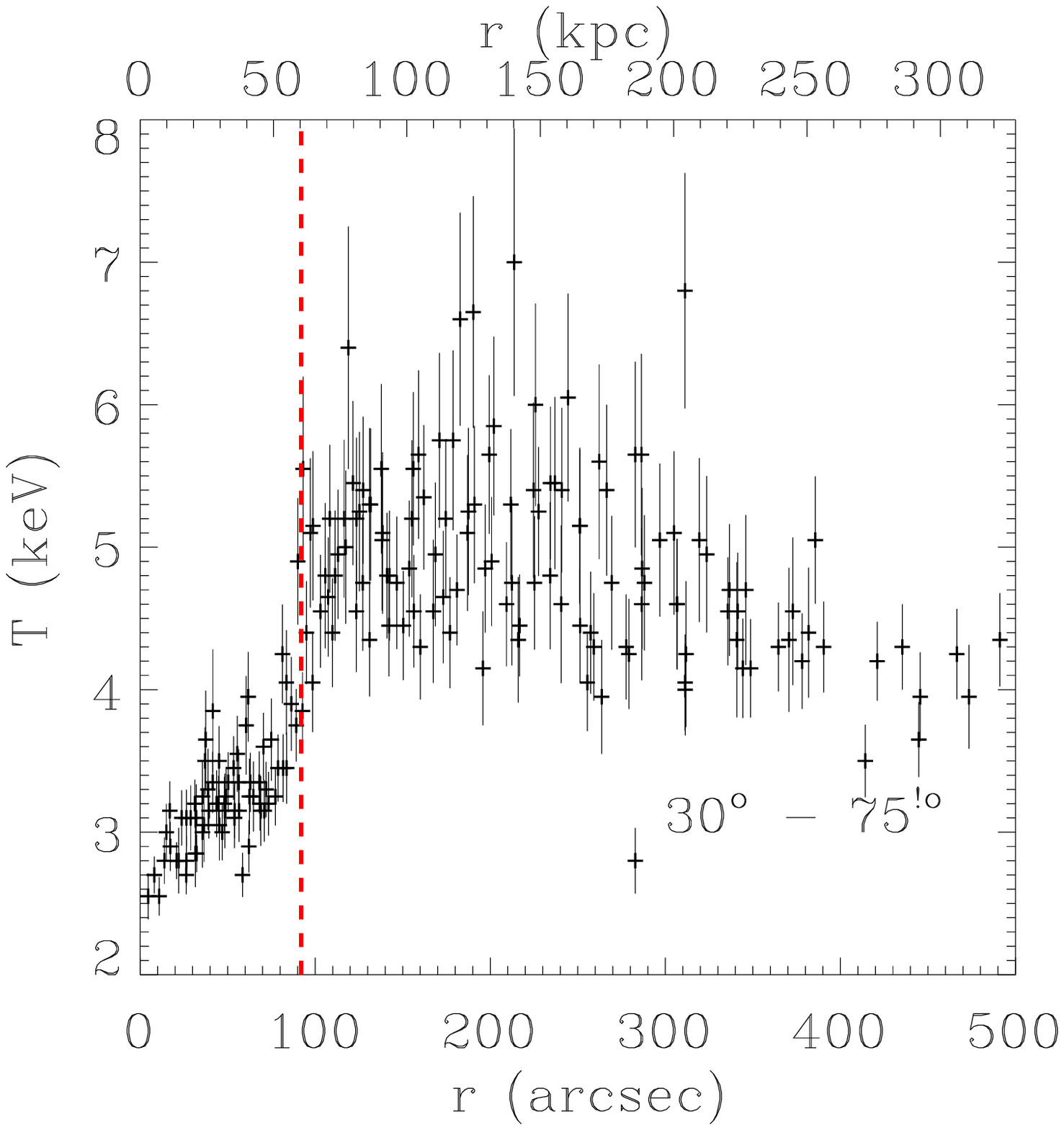}}
\vspace{-2truemm}
\centering
{{\includegraphics[width=0.25\textwidth]{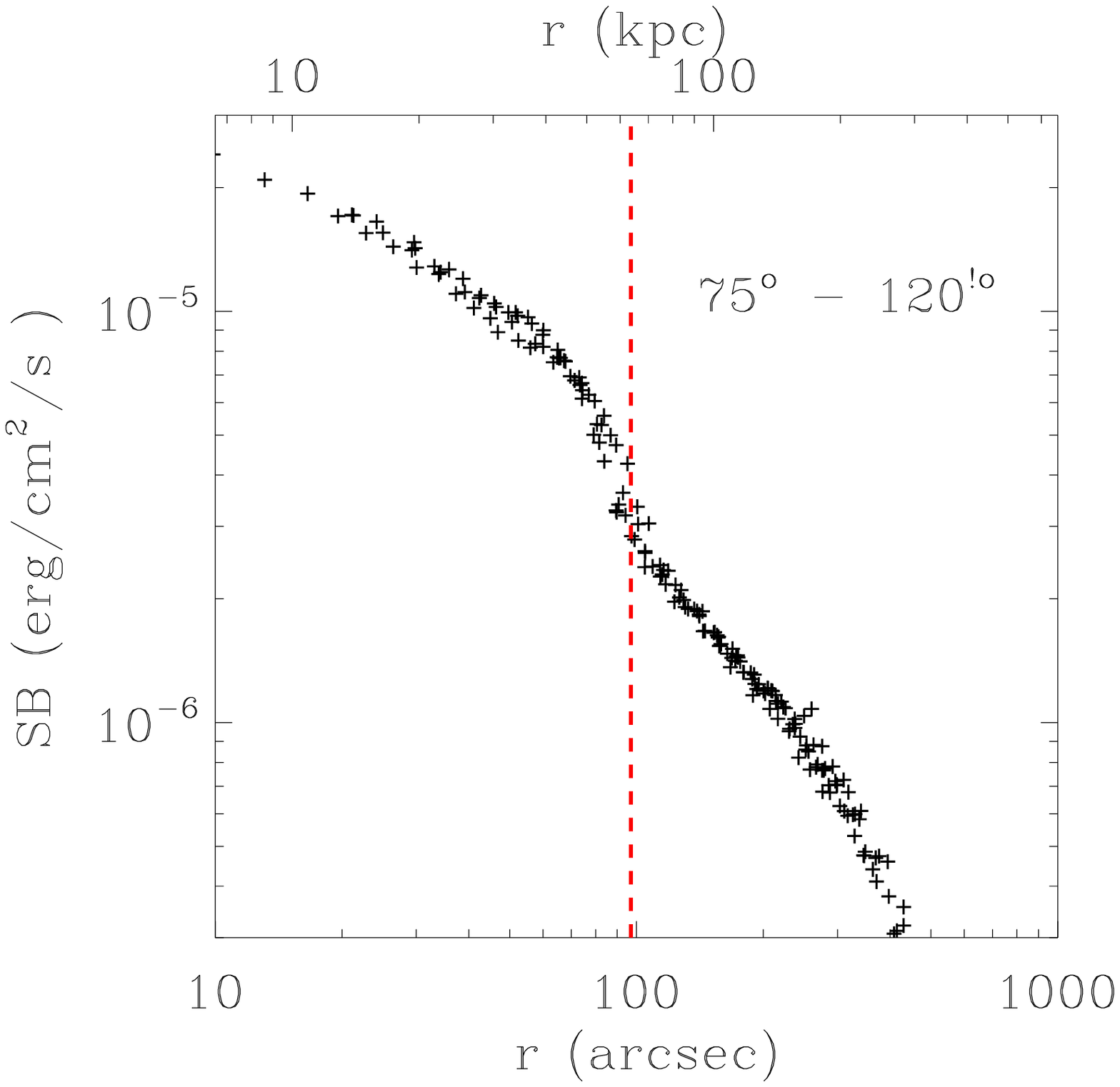}}
\hspace{-7truemm} 
{\includegraphics[width=0.25\textwidth]{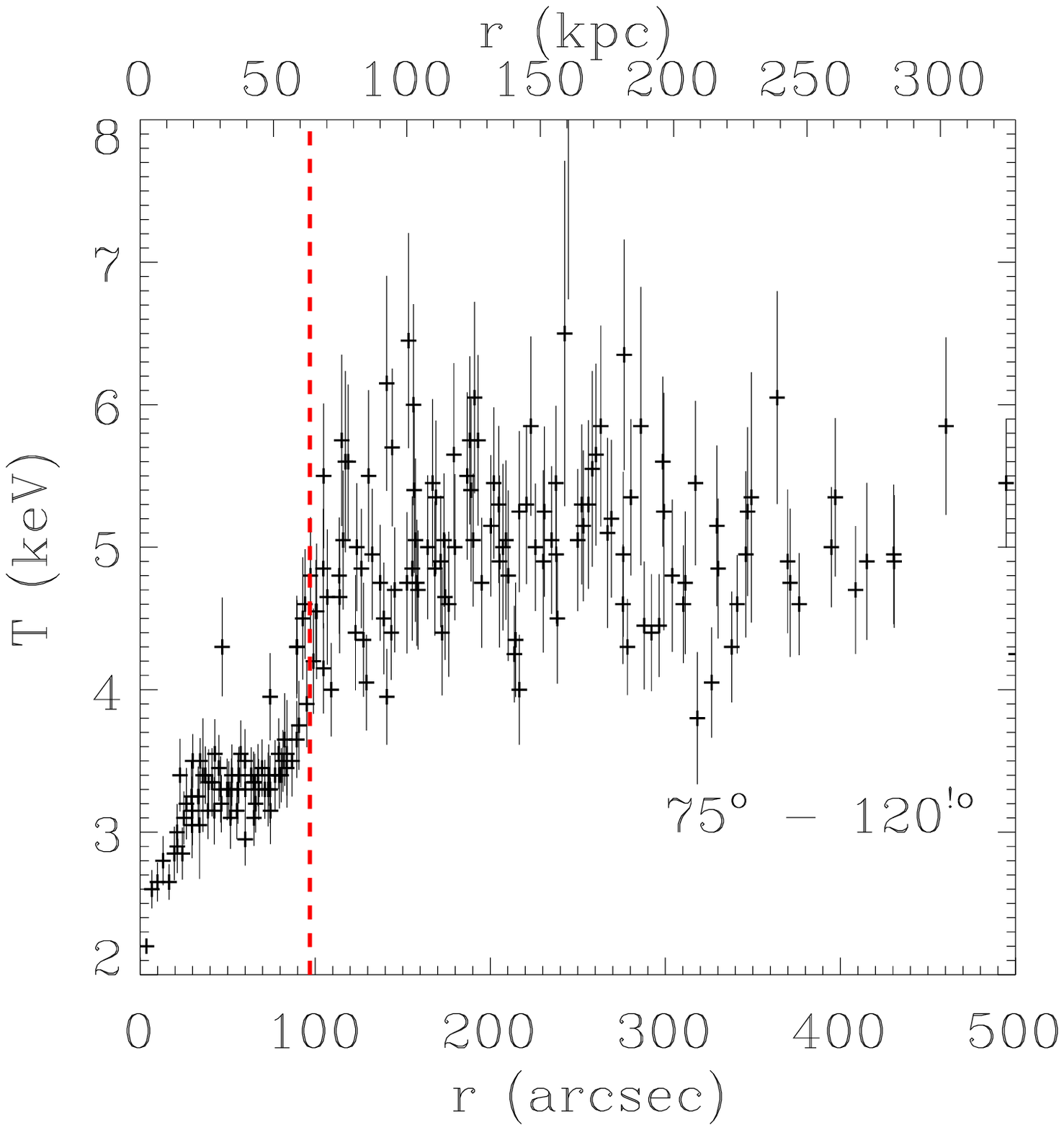}}}
\centering
{{\includegraphics[width=0.25\textwidth]{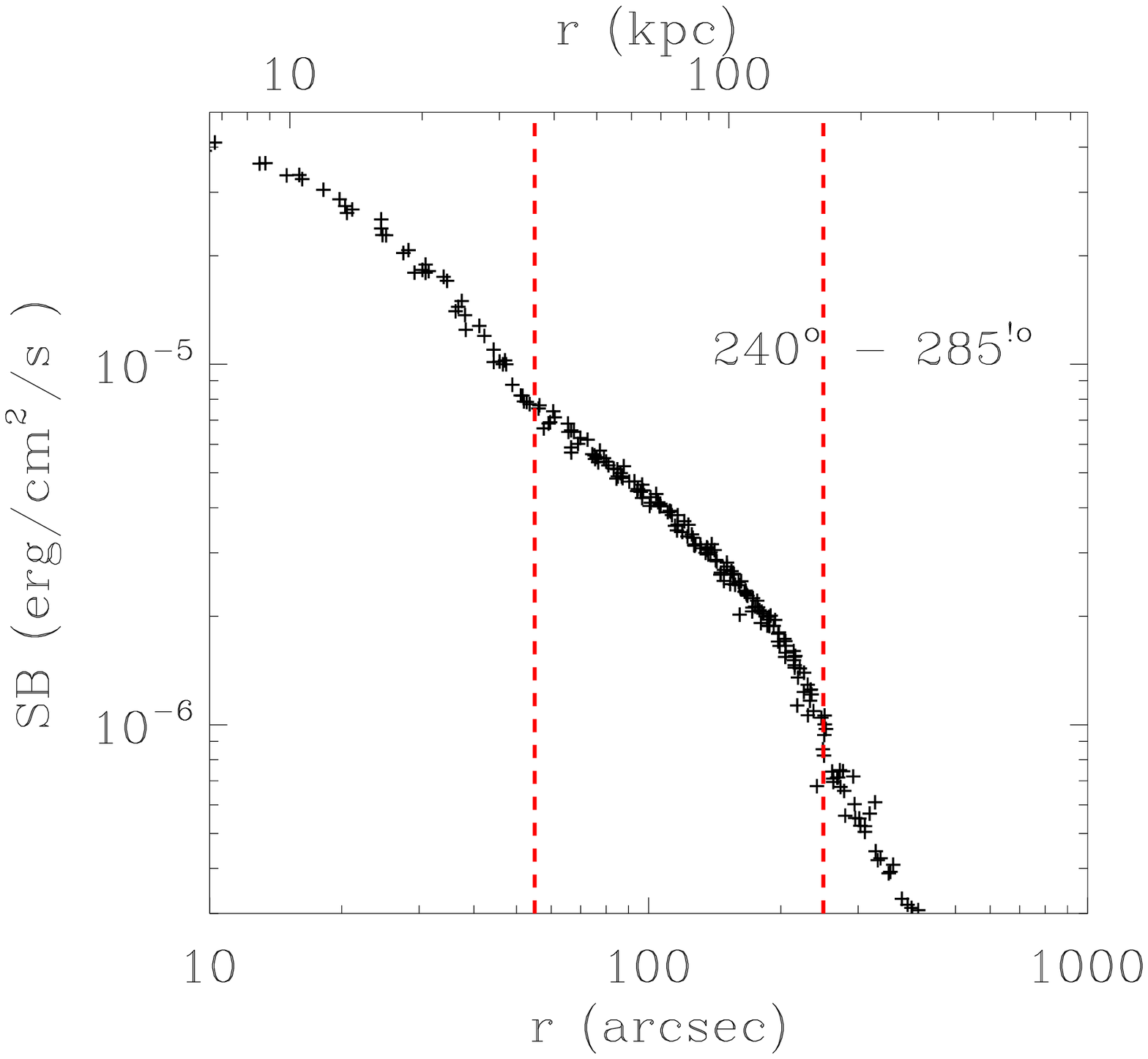}}
\hspace{-7truemm}
{\includegraphics[width=0.25\textwidth]{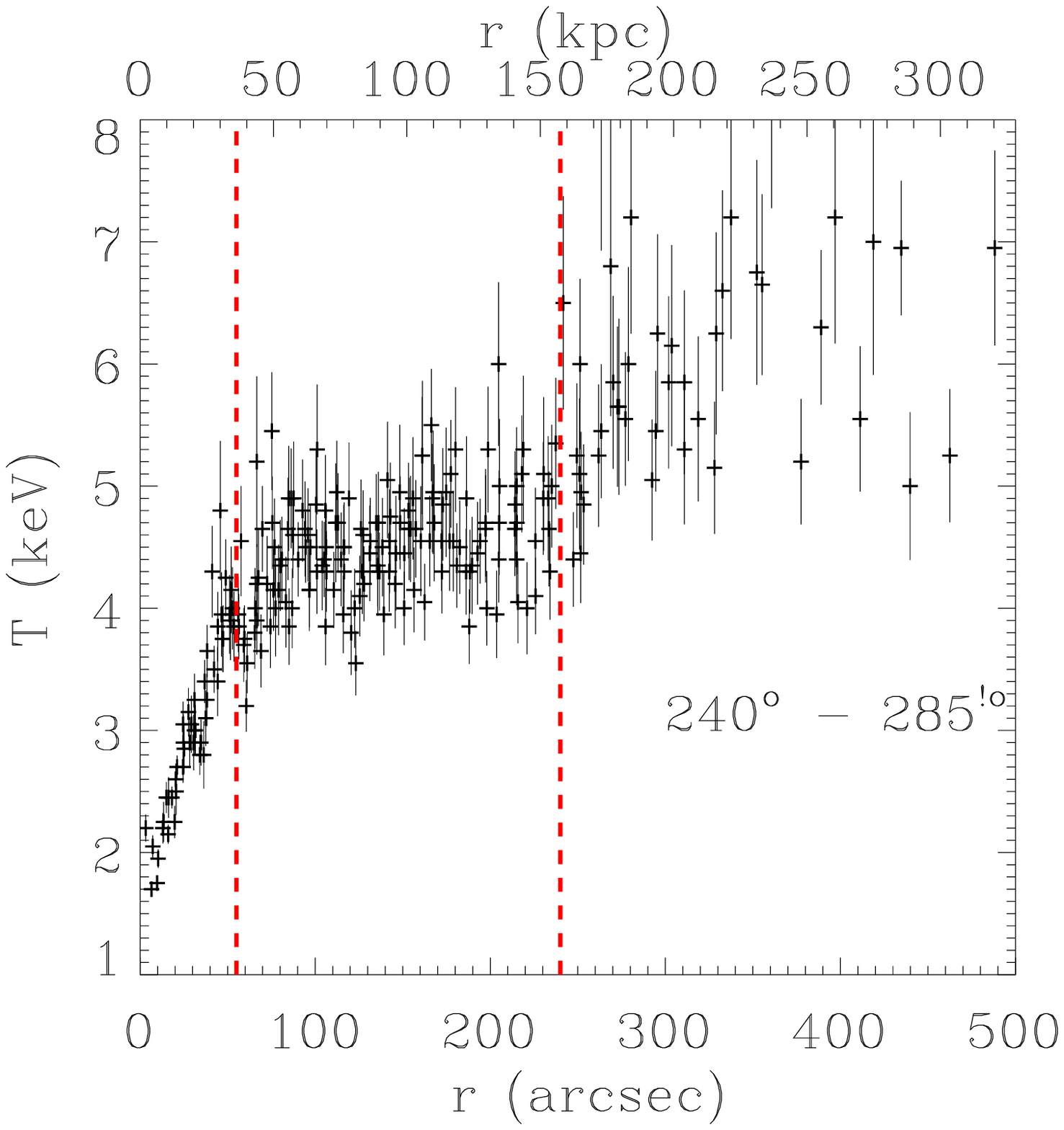}}}
\caption{Surface brightness and temperature profiles obtained from 
maps of Fig \protect\ref{fig:mappe_wvt} for the sectors hosting the cold fronts. Red vertical lines mark 
the cold front positions.}
\label{fig:SB_T_profs}
\end{figure}

The temperature map (panel (b) in Fig. \ref{fig:mappe_wvt}) is typical of a cool core cluster: 
the outer temperature is about 5 keV in the external regions and it decreases 
when moving towards the center down to $\sim 2$ keV.
The map shows that temperature rises are associated to the surface brightness discontinuities as expected for cold fronts.
To quantify these discontinuities, we extract the surface brightness and temperature profiles in the sectors hosting the cold fronts. 
We split the 30$^o$-120$^o$ sector 
hosting the NNW main discontinuity into two sub-sectors (30$^o$-75$^o$ and 75$^o$-120$^o$) because 
of the boxy morphology of this feature.
The profiles are plotted in Fig. \ref{fig:SB_T_profs}: the vertical red lines mark the cold fronts positions in each sector. 
Notably, the 240$^o$-285$^o$ sector hosts two discontinuities. The plots confirm that the temperature increases across all 
the discontinuities detected and that all these discontinuities 
(four, when including also the E cold front detected by Dupke \& White~2003) are definitively cold fronts. 

The temperature map shows that the central cool gas develops in a spiral-shaped pattern extending from the center 
anti-clockwise towards north (where the main NNW cold front is located); then the spiral turns towards East and finally 
expands in the South direction. 
The cold fronts correspond to the edge of the spiral and the southern outermost (S2) cold front 
is located at the tail of the spiral. 

The entropy (panel (d) in Fig. \ref{fig:mappe_wvt})
follows the temperature behaviour and, as expected in a sloshing scenario (G10) it decreases rapidly 
in the central core.
The pressure (panel (c) in Fig. \ref{fig:mappe_wvt}) is nearly symmetric, suggesting that the cluster is 
fairly relaxed and that no remarkable perturbations to the underlying gravitational potential
have been induced  by any major merger event, as expected in a sloshing scenario,
where the mechanism is triggered by minor mergers.

\subsection{Metal distribution}

\begin{figure}
\vspace{-3truemm}
{\includegraphics[width=0.45\textwidth]{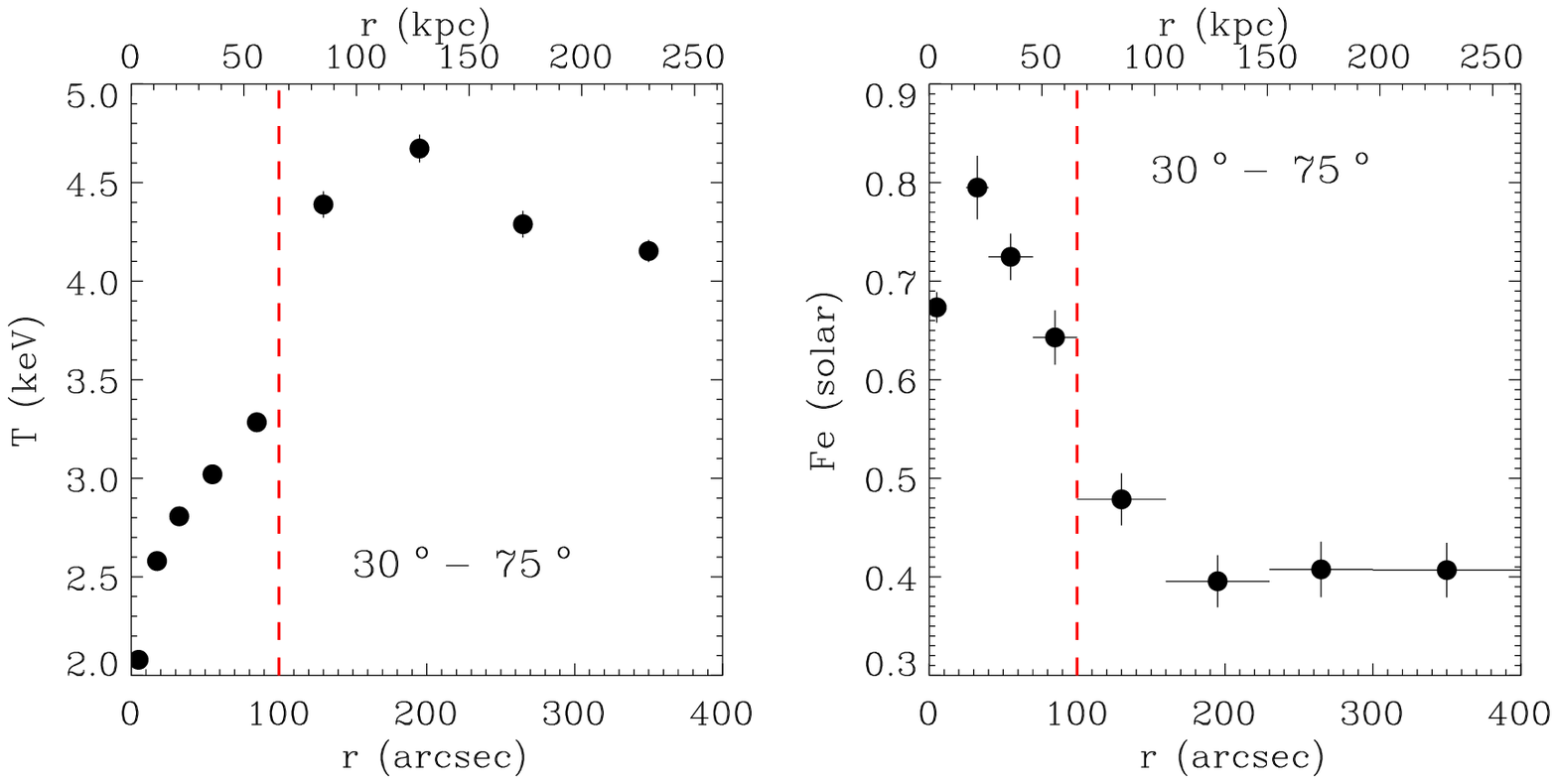}}
\vspace{-3truemm}
{\includegraphics[width=0.45\textwidth]{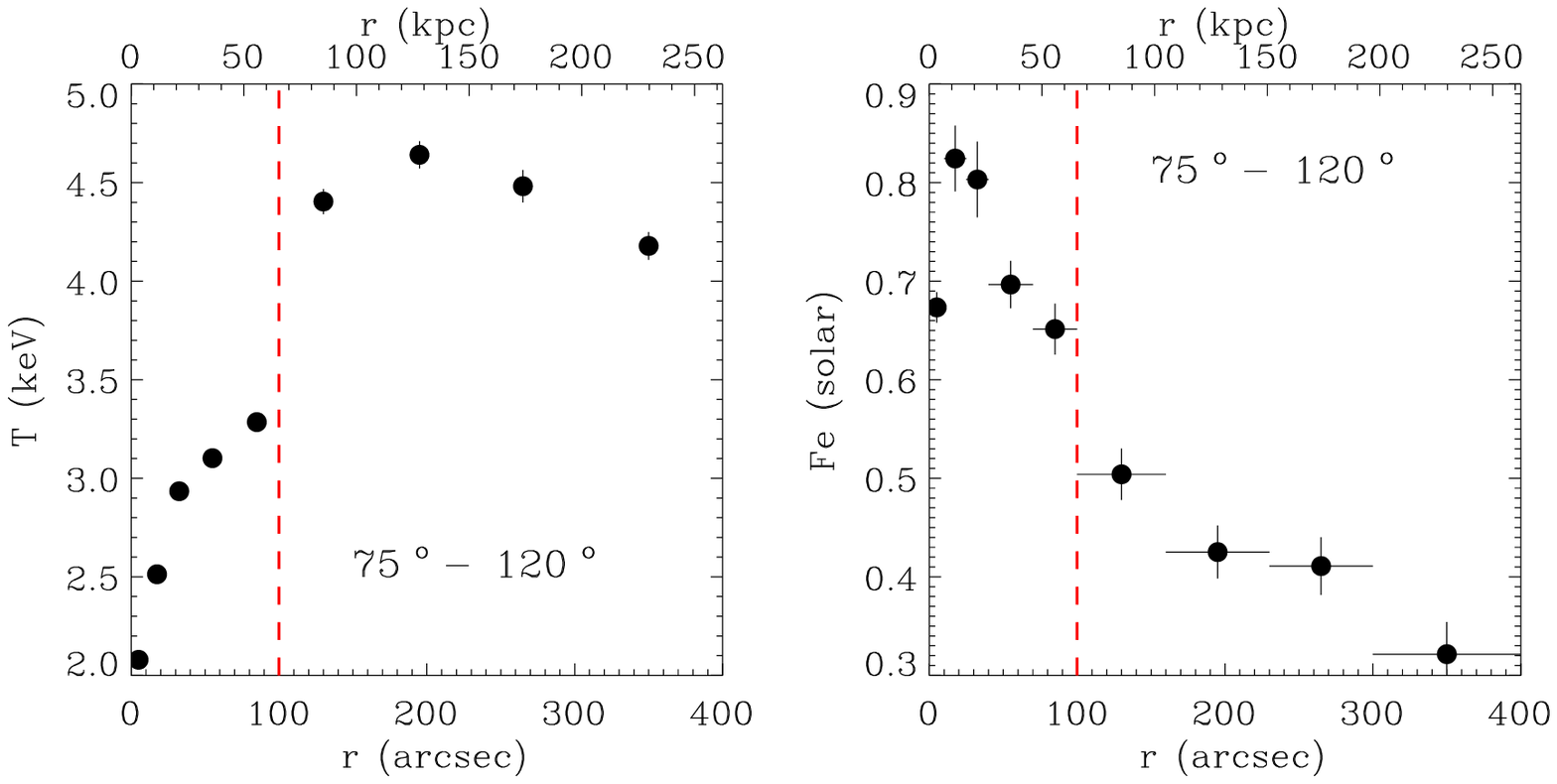}}
{\includegraphics[width=0.45\textwidth]{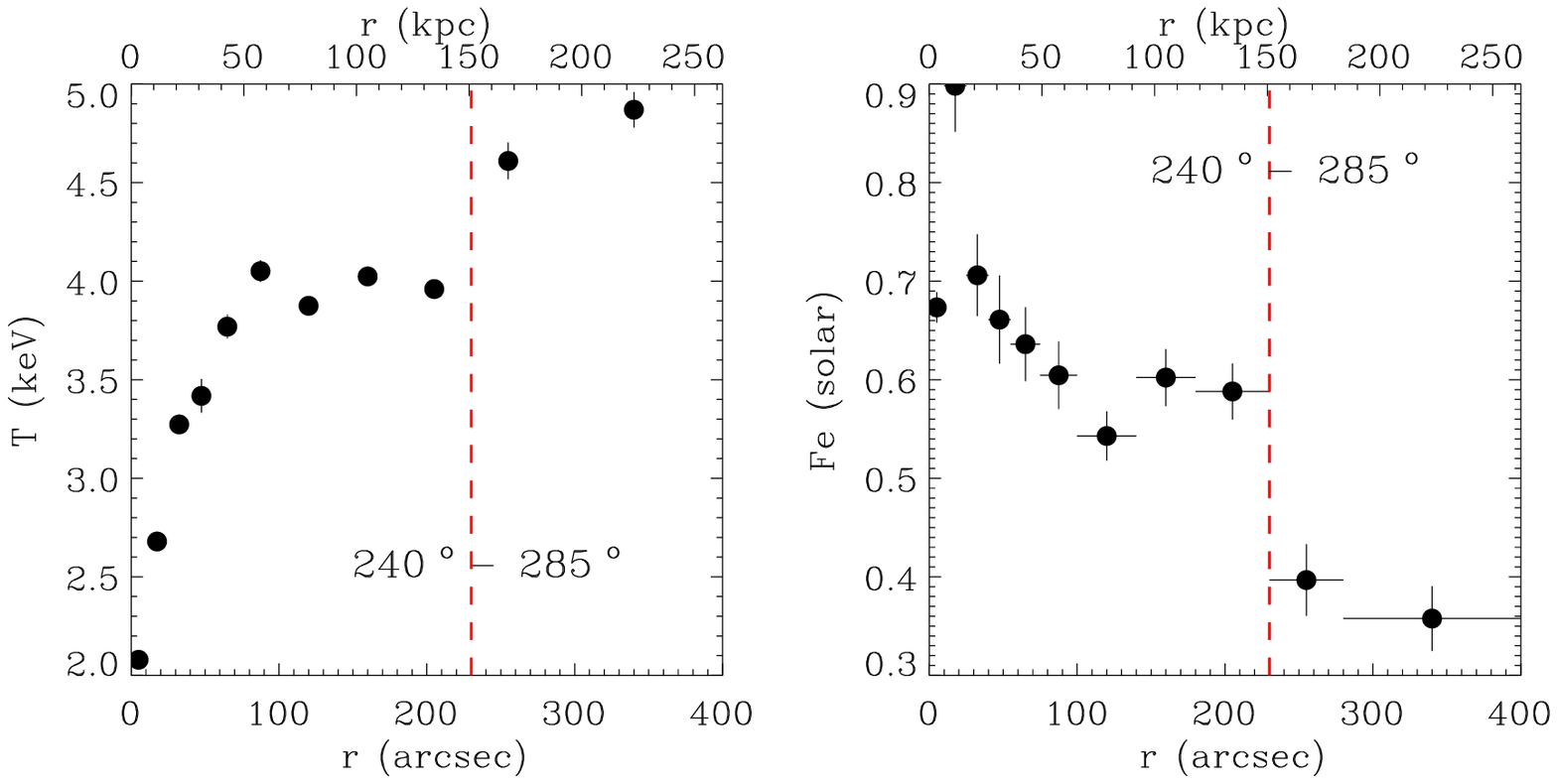}}
\caption{Temperature (panels on the left) and metal abundance (panels on the right) profiles from spectral 
analysis for the sectors hosting cold fronts. Red vertical lines mark the cold front positions.}
\label{fig:zprofs}
\end{figure}

\begin{figure}
\centering
 \includegraphics[width=70mm,angle=90]{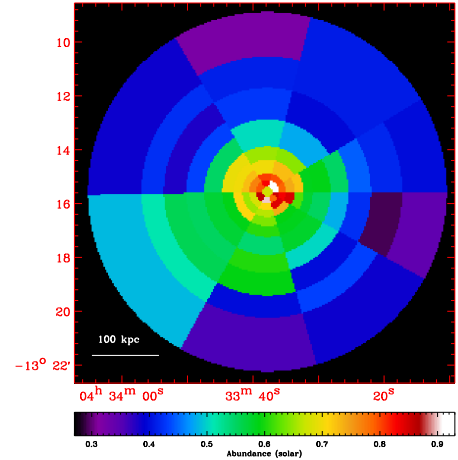}

\caption{Metal abundance map for A496.}
\label{fig:zmap_annuli}
\end{figure}

In order to study the distribution of metals in A496, we divide the cluster into rings and sectors. 
Both are selected to match the cold fronts positions. 
For each selected region, we extracted the spectra for the three EPIC detectors and fitted them simultaneously with a 
single temperature {\it apec} plasma model using the {\it xspec} software.
The metallicity map obtained from this spectral analysis is shown in Fig. \ref{fig:zmap_annuli} and  
we plot the temperature and abundance profiles 
for the sectors of interest in Fig. \ref{fig:zprofs}.
We focus on the main (NNW) cold front and on the outermost southern S2 cold front.
Here, the metallicity exhibits the typical behavior of a cool core cluster, with a central abundance peak 
($\sim$0.9 solar) and a decreasing trend towards the outskirts where it reaches the value of $\sim$0.3-0.4 solar.
In the central bin a significantly lower value is measured. This is due to the iron-bias and the feature disappears 
when using a two temperature model for the central bin (see Ghizzardi at al., in preparation).

The metal abundance drops abruptly across the S2 cold front (240$^o$-285$^o$), 
similarly to what observed in other cool core clusters (Perseus: Fabian et al.~2011; Centaurus: Sanders \& Fabian~2006; 
A2204: Sanders, Fabian \& Taylor~2005, 2009). 
A significant decrease of the metal abundance, consistent with a discontinuity, is also 
observed across the NNW cold fronts (30$^o$-75$^o$ and 75$^o$-120$^o$) where we detect the most significant variation of 
metallicity: the profile gradient across the cold front is higher than in any other region of the profile.

In the sector 240$^o$-285$^o$, the metallicity of the area inside the outermost cold front (radius in the range 60-150 kpc) 
keeps constant on a plateau at $\sim$0.6 solar, before dropping at the cold front position.
This can be better seen in the metallicity map (Fig. \ref{fig:zmap_annuli}).
The map shows that the external metallicity is almost constant at a value of 0.3-0.4 solar 
in all the directions except for the sector 180$^o$-285$^o$ where an excess of metals ($\sim$0.6 solar)
can be clearly observed. This sector corresponds to the region of the spiral tail.

\section{Discussion and summary}

We analyzed a long XMM-Newton observation of A496. Taking advantage of the XMM-Newton large collecting area and 
spectral resolution  the quality of the data is good enough to study in detail the metal abundance for this cluster 
and the correlation of the metal distribution with cold front positions. 
A496 is particularly suitable to investigate this issue: our analysis (in agreement with previous results) shows 
that this cool core cluster hosts at least 4 cold fronts.
The temperature map exhibits a spiral-like pattern of cool gas which 
wraps around the center anti-clockwise. The presence of multiple cold fronts and of the spiral feature are 
the typical signatures of sloshing of the central cool gas.
The spiral morphology is characteristic of a sloshing scenario as the central 
cool gas may acquire angular momentum during an off-axis minor merger (Ascasibar \& Markevitch~2006).

We performed a spectral analysis to build metal abundance profiles and we find that the metallicity has a discontinuity 
across all the cold fronts. According to this, a general picture can be drawn for the metal behavior during the 
sloshing evolution: when the sloshing starts  the innermost cold, dense and metal rich gas 
is displaced from the center into a hotter and less abundant region of the cluster. 
This displacement creates the cold front feature characterized by the density and temperature discontinuities, and 
also the observed metal jump across the front. As the colder gas sloshes towards the potential well center, 
it carries the metals keeping its high metallicity, hence the metal jump across the front is preserved.

We also find a metal excess in the region hosting the tail of the spiral-like structure. 
This suggests that the cool central gas drags the heavy elements while swirling around the center
of the potential well and that the  sloshing is able to move metal from the center in the outer 
part of the cluster as the spiral expands towards the outermost regions of the cluster. 



\end{document}